\documentclass[aps,prb,twocolumn,letterpaper]{revtex4}
\usepackage{graphicx}
\usepackage{dcolumn}
\usepackage{amsmath}
\usepackage{amssymb}
\usepackage[caption=false]{subfig}
\usepackage{soul}
\usepackage{hyperref}

\newcommand{\subfigimg}[3][,]{%
  \setbox1=\hbox{\includegraphics[#1]{#3}}% Store image in box
  \leavevmode\rlap{\usebox1}% Print image
  \rlap{\hspace*{0pt}\raisebox{\dimexpr\ht1-1.5\baselineskip}{#2}}% Print label
  \phantom{\usebox1}% Insert appropriate spcing
}

\usepackage[usenames,dvipsnames]{color}
\definecolor{orange}{rgb}{1,0.5,0}
\usepackage{float}
\usepackage{epstopdf}
\usepackage{ulem}
\def\vv{{\bf v}}

\def\kv{{\bf k}}

\def\beq{\begin{equation}}
\def\eeq{\end{equation}}

\def\beqa{\begin{eqnarray}}
\def\eeqa{\end{eqnarray}}

\begin{document}

\title{Quantum oscillations in Weyl semimetals - a surface theory approach}
\author{Jan Borchmann and T. Pereg-Barnea}
\affiliation{Department of Physics and the Centre for Physics of Materials, McGill University, Montreal, Quebec,
Canada H3A 2T8}
\date{\today}
\begin{abstract}
We develop an effective surface theory for the surface states of a Weyl semimetal.  This theory includes the peculiar Fermi arc states on the surface as well as leakage of the states from the surface to the bulk.  Subjecting the model to a magnetic field perpendicular to the surface results in quantum oscillations. The oscillations are different from the usual ones since they do not involve a closed Fermi surface cross section.  It has been shown previously that the Quantum oscillations can be understood semiclassically as resulting from motion of electrons on the surface Fermi arcs as well as tunneling through chiral Landau levels associated with the bulk.  In this work we develop an effective surface theory and use it to analyze the quantum oscillation in the semiclassical regime and beyond.  Specifically, we show that when a pair of Weyl points are close to each other the surface quantum oscillations acquire a phase offset which originates from the bulk. While the surface states are responsible for a large part of the electron motion, tunneling through the bulk is necessary for completing the orbit.  This tunneling makes use of the bulk, zero energy, chiral Landau level in each Weyl node.  When the nodes are close in momentum space their chiral levels overlap and a gap at zero energy is formed.  This gap causes the phase offset in the surface quantum oscillations. 
\end{abstract}
\maketitle
\section{Introduction}
Topological semimetalic states of matter have been theoretically predicted in 2007 by Murakami\cite{murakami} and a simple lattice model of a Weyl semimetal (WSM) was proposed by Burkov and Balents in 2011\cite{burkov}.  The early studies led to further theoretical work\cite{wan,burkov2,witczak,chen2,heikkil} as well as the experimental realization of a type-I WSM in TaAs\cite{weng,huang,lv,lv2,yang2,belopolski}, TaNb\cite{xu3}, NbP\cite{liu} and TaP\cite{xu4}. Type-II WSM behavior has been predicted in $\text{WTe}_2$\cite{soluyanov}, $\text{TaIrTe}_4$\cite{koepernik} as well as $\text{MoTe}_2$\cite{sun, wang}. Additionally, several theoretical predictions of Weyl semimetals have been made including in  $\text{SrSi}_2$\cite{huang2}, $\text{HgCr}_2\text{Se}_4$\cite{xu} and $\text{Mo}_x \text{W}_{1-x}\text{Te}_2$\cite{chang2}. 

Weyl semimetals are three dimensional materials which are characterized by band crossing points in momentum space.  Close to these points the energy disperses linearly with momentum and the system is described by the Weyl Hamiltonian.  In three dimensions the Weyl nodes are robust against deformations and appear in pairs due to time reversal or inversion symmetry.
The Weyl points are Berry curvature monopoles in momentum space and are responsible for the chiral anomaly\cite{nielsen,adler,bell} which is exhibited in a variety of transport effects\cite{hosur} such as negative magnetoresistance\cite{nielsen}, the anomalous quantum Hall effect\cite{burkov2,xu,yang,zyuzin,vafek,goswami,chen}, the chiral magnetic effect\cite{zhou,vazifeh,chen} as well as coupling between magnons and plasmons\cite{liu}.

When the Fermi level of the system is close to the energy of the Weyl nodes the bulk low energy properties of the system are governed by these points.  Expanding the Hamiltonian to linear order in momentum about the Weyl points leads to,
\begin{align}
\begin{split}
H_W = \hbar \vv\cdot \kv \ \sigma_0 + \sum_{i,j=x,y,z} \hbar h_{ij}k_i\sigma_j, \label{h_weyl}
\end{split}
\end{align}
where the indices $i$ and $j$ run over spatial directions, $\sigma_0$ is a two dimensional unit matrix while other $\sigma_n$s are Pauli matrices. The second term in the Hamiltonian is the usual Weyl Hamiltonian in which the matrix $h_{ij}$ determines the spinor direction.  The determinant of $h_{ij}$ is $\chi=\pm1$, the chirality of the Weyl point.  The first term is unique to condensed matter systems, as it breaks Lorentz invariance. Unlike the second term, it is proportional to the unit matrix.  This term, named 'tilt' does not influence the spin direction but does have an effect on the spectrum as it tilts the Weyl cone in the energy-momentum space.
In type-I WSM the tilt term is not strong enough to alter the nature of the Weyl point while in type-II WSM the tilt causes the system to have a finite density of states at the Weyl node energy.  The density of states at zero energy alters the low temperature thermodynamic properties such that type I and type II WSMs can be distinguished experimentally\cite{soluyanov}.

Pairs of Weyl nodes with opposite chirality and at different momentum also lead to Fermi arcs.  These are zero energy states localized on the sample surface and characterized by surface momentum along a line connecting the Weyl nodes projected to the surface momentum.  Their existence can be understood in the context of the quantum anomalous Hall effect in two dimensions by slicing the system into many two dimensional systems.  In each slice a topological invariant can be calculated and the presence of the Weyl nodes indicate a jump in this number.  The Fermi arc states are therefore the collection of zero energy chiral edge states.\cite{burkov} In this paper we concern ourselves with studying the Fermi arc states and their response to magnetic field. 

In order to address the surface states we define the surface Brillouin zon (SBZ) which is made of the allowed momenta parallel to the surface.  The behaviour of single particle wavefunctions perpendicular to the surface is described in real space. As mentioned above, the zero-energy surface states form Fermi arcs\cite{wan} which have been observed in photoemission measurement in TaAs\cite{xu2}. 
Recently, Potter {\it et al.}\cite{potter,zhang} proposed that the surface states on the arcs respond to a perpendicular magnetic field and produce magnetic quantum oscillations. 
Semi-classically the electrons on the Fermi arcs on the two surfaces connect though the bulk chiral bulk Landau level and can therefore move on a closed magnetic path.  The semiclassical analysis predicts the following level quantization:\cite{potter}
\begin{align}
\begin{split}
\epsilon_n = \frac{\pi\hbar v}{k_a \ell_B^2 + L} \left( n + \gamma \right), \label{arcLL}
\end{split}
\end{align}
where, $k_a$ is the length of the Fermi arc, $l_B$ is the magnetic length, $L$ is the thickness of the slab and $v$ is the Fermi velocity.  The phase shift $\gamma$ not accessible semiclassically.   The period of the oscillations has two contributions. The first is given by the propagation of the electrons along the arc and is therefore proportional to the arc length $k_a$. The second contribution is proportional to $L$ since it is due to tunneling through the bulk. This dependence of the quantum oscillations on the geometry of the sample provides a signature of the mixed surface-bulk magnetic path, unique to WSM.  Experimental evidence of this dependence was found in $\text{Cd}_3\text{As}_2$\cite{moll}.

This rest of this paper is structured as follows. In section~\ref{Sec:effective} we introduce the model and develop the effective surface theory. In section~\ref{Sec:field} we apply a magnetic field to the system and use the effective surface theory to study the surface quantum oscillations and study the short arc length regime. In section~\ref{Sec:numeric} we compare our findings to a full numerical treatment of the WSM slab. 
%Our conclusions are given in~\ref{Sec:conclusion}.

\section{Effective Surface Theory}\label{Sec:effective}
\subsection{The model}
We begin with a two-orbital tight binding model\cite{chang},
\begin{align}
\begin{split}
H_{3D} ={}& t_{s}\left( \sin{k_x}\sigma_x + \sin{k_y}\sigma_y + \sin{k_z}\sigma_z \right) \\
 &+ \left( m + t^{\prime}(2 - \cos{k_x} - \cos{k_y}) \right)\sigma_z,
\end{split}
\end{align}
where time-reversal symmetry $H(\mathbf{k}) = \sigma_y H^*(-\mathbf{k})\sigma_y$ is broken by the second term. 
The choice to break time reversal symmetry is not limiting as one could devise a similar, time reversal invariant Hamiltonian, while breaking inversion symmetry.   Our choice here is made in order to work with small matrices.
The bulk energies are given by,
\begin{align}
\begin{split}
E_\pm = & \pm \left[ t_s^2 \left( \sin^2{k_x} + \sin^2{k_y}\right) \right. \\ 
& + \left. \left( t_s \sin{k_z} + m + t^{\prime}(2 - \cos{k_x} - \cos{k_y}) \right)^2  \right]^{\frac{1}{2}}. \label{dispersion}
\end{split}
\end{align}
Throughout the remainder of the paper, we will set $t^\prime = 1$ and measure energy in units of $t^\prime$. The nodes of Eq.~(\ref{dispersion}) show that the model has different phases depending on the parameters $m$ and $t_s$. For example, when keeping $t_s$ fixed to 1 and varying $m$, one finds the following phases:
\begin{itemize}
\item For $m> t_s$ the model is gapped and trivial.
\item At $m=t_s$ a gap closure appears at $\mathbf{k} = (0,0,-\pi/2)$.
\item For $-t_s<m<t_s$ the gap closure splits into two Weyl nodes which recombine for $m=-t_s$ at $\mathbf{k} = (0,0,\pi/2)$.
\item For $-3t_s<m<-t_s$ there are two pairs of Weyl nodes which appear for $m=-t_s$ at $\mathbf{k} = (\pi,0,-\pi/2)$ and $\mathbf{k} = (0,\pi,-\pi/2)$. 
\item The Weyl points recombine again at $m=-3t_s$ at $\mathbf{k} = (\pi,0,\pi/2)$ and $\mathbf{k} = (0,\pi,\pi/2)$.
\item For $-5t_s<m<-3t_s$ two Weyl nodes emerge at $\mathbf{k} = (\pi,\pi,-\pi/2)$ when $m=-3t_s$ and split when $m$ is decreased.  The two points recombine at $\mathbf{k}  = (\pi,\pi,\pi/2)$ when $m=-5t_s$.
\item For $m<-5t_s$ the model is again gapped and trivial.
\end{itemize}
We choose to work in one of the Weyl semimetal regimes above, where $|m|<t_s$. The analysis can be easily extended to other regimes.

We work in a slab geometry with $(010)$ surfaces such that $k_x$ and $k_z$ remain good quantum numbers. We work partially Fourier transformed operators, $c_{i_y,\vec{k}}$, where $\vec k=(k_x,k_z)$ and $i_y$ is a discrete coordinate in the $\hat y$ direction ranging between $1$ and the number of layers, $N_y$.  With this definition the Hamiltonian can be written as a block matrix in $i_y$ and spin space.  The diagonal blocks $H_0(\vec k)$ represent hopping within the $x-z$ layer while the off diagonal ones represent inter-layer terms.  The inter-layer terms connect neighbouring layers $i_y$ and $i_y\pm1$ and are given by $2\times 2$ matrices in spin space, $R$.  We can therefore write the system's three dimensional Hamiltonian matrix as:
\begin{align}
\begin{split}
H_{3D} = \begin{pmatrix}
H_0 & R & 0 & 0 & \dots & 0 & 0 \\
R^\dagger & H_0 & R & 0 & \dots & 0 & 0 \\
0 & R^\dagger & H_0 & R & \dots & 0 & 0 \\
\vdots & \vdots & \vdots & \vdots & \ddots & \vdots & \vdots \\
0 & 0 & 0 & 0 & \dots & R^\dagger & H_0 \\
\end{pmatrix},
\end{split}\label{eq:H3D}
\end{align}
with
\begin{align}
\begin{split}
R =  \begin{pmatrix}
-\frac{1}{2} & -\frac{t_s}{2} \\
\frac{t_s}{2} & \frac{1}{2} \\
\end{pmatrix}.
\end{split}
\end{align}
and
\begin{align}
\begin{split}
H_0 &= t_{s} \sin{k_x}\sigma_x + \left( m + 2 - \cos{k_x} + t_s \sin{k_z} \right)\sigma_z \\ 
&\equiv g_1(\vec k)\sigma_x + g_3(\vec k)\sigma_z.\label{model}
\end{split}
\end{align}
In this geometry, surface states appear and are arranged in arcs in the surface Brillouin zone.  The arcs extend between the projections of the Weyl points onto the surface Brillouin zone. In our lattice model and in the regime we choose to work, the arc connecting the two Weyl nodes is a straight line along the $k_z$-axis of length $k_a = 2\arccos{(\frac{m}{t_s})}$.
\subsection{Effective surface propagator}
In order to derive an exact effective surface theory, we treat the surface degrees of freedom independently from the bulk and integrate out the bulk degrees of freedom. We denote the sites with $i_y = 1,N_y$ as 'surface' and the sites with $1<i_y<N_y$ as bulk.  The matrix $H_{3D}$ is rearranged in this manner and one can identify 4 blocks: $H_b$ is an $2(N-2)\times2(N-2)$ matrix which contain the bulk terms, $H_s$ is a $4\times4$ matrix containing terms within the two surfaces and the off diagonal blocks $T$ and $T^\dagger$ couple the two. They are $2(N-2)\times 4$ and $4\times 2(N-2)$ matrices.
Following Marchand and Franz\cite{marchand}, the expression for the surface Green's function, 
\begin{equation}
G_{\text{eff}} (i\omega_n) = \left[ G_s^{-1}(i\omega_n) - T^\dagger G_b (i\omega_n)T \right]^{-1},\label{geff}
\end{equation}
where $G_{b,s}(i\omega_n) = -(i\omega_n - H_{b,s})^{-1}$ are the uncoupled bulk and surface Greens functions, respectively. It is important to note that the propagator in Eq.~\ref{geff} is not directly related to an effective Hamiltonian since it contains a finite lifetime due to the decay of surface states into the bulk.

\subsection{Numeric evaluation of the effective Green's function in a slab geometry}
The Hamiltonian of the full three dimensional system was given in Eq.~\ref{eq:H3D}.  In a slab geometry the bulk Hamiltonian is the same as the full Hamiltonian with the first and last two rows and columns removed.  The surface Hamiltonian reads:
\begin{eqnarray}
H_s &= \begin{pmatrix}
H_0 & 0\\
0 & H_0
\end{pmatrix},
\end{eqnarray}
and the coupling matrix is given by:
\begin{eqnarray}
T =& \begin{pmatrix}
R^\dagger & 0 \\
0 & 0 \\
\vdots & \vdots \\
0 & 0 \\
0 & R \\
\end{pmatrix},
\end{eqnarray}
with $N-2$ such rows.

\begin{figure}[t]
\begin{center}
\subfloat{\subfigimg[width=0.23\textwidth]{(a)}{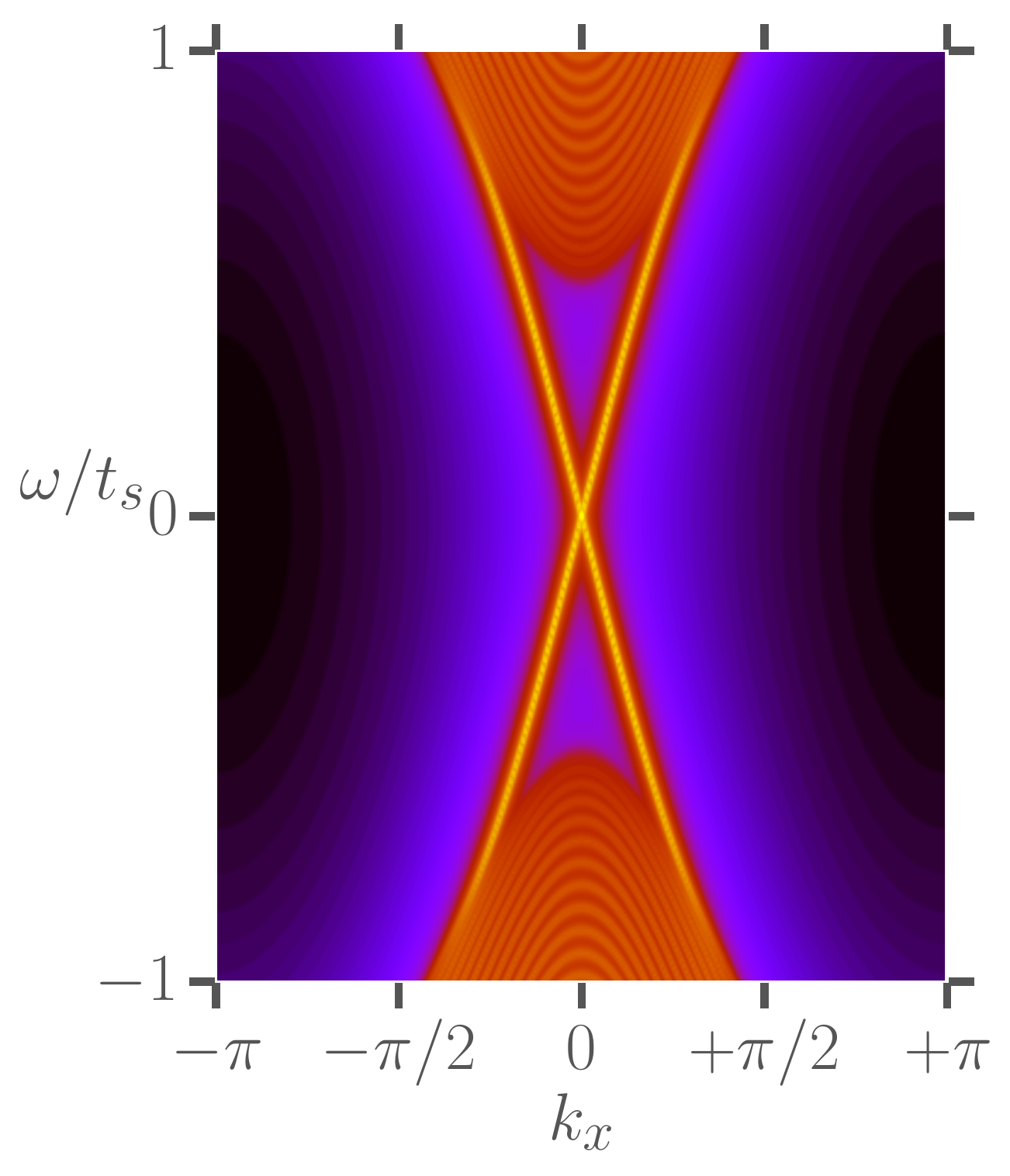}}\\[-0.3em]%\hspace*{-0.2em}
\subfloat{\subfigimg[width=0.23\textwidth]{(b)}{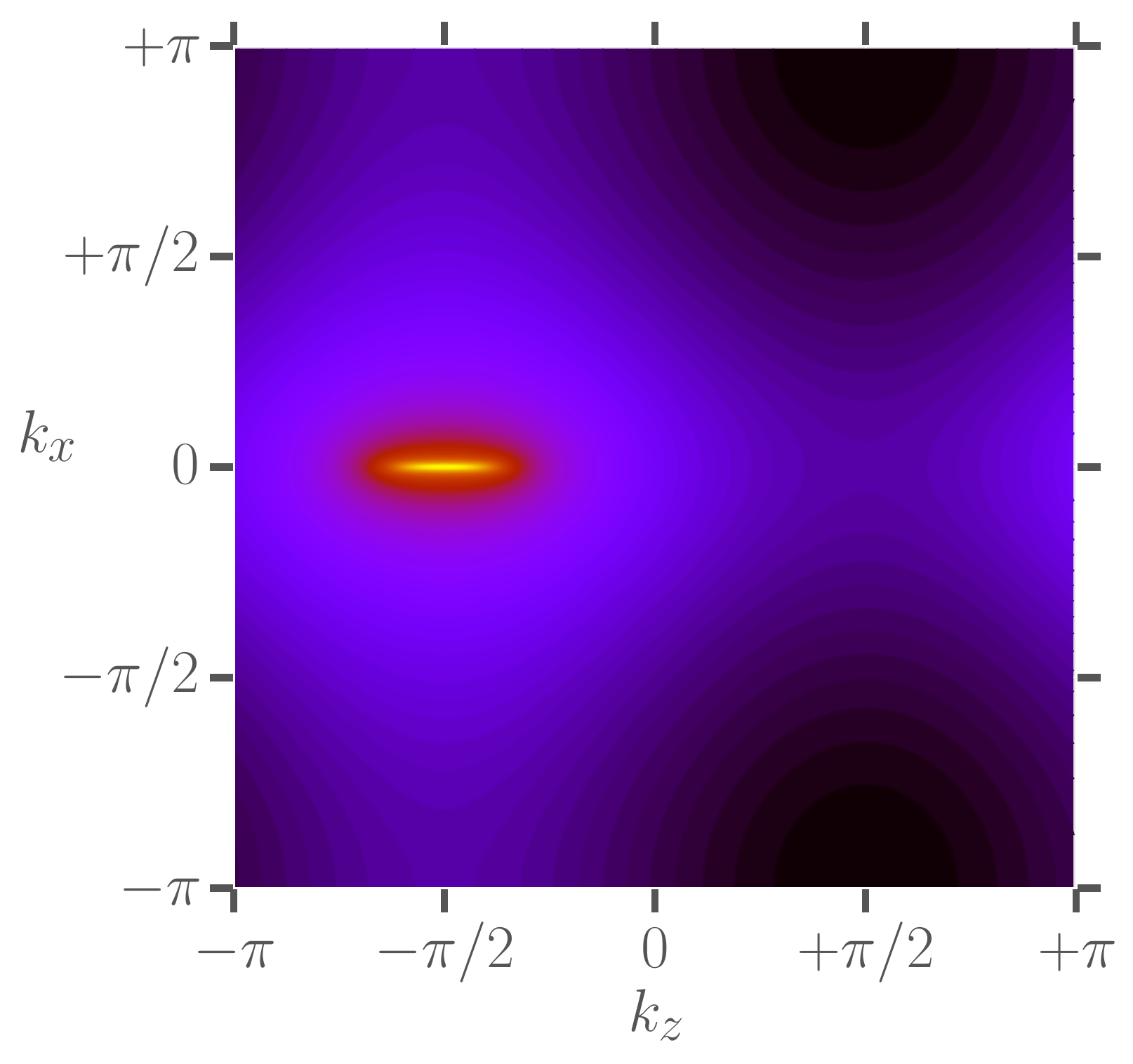}}
\subfloat{\subfigimg[width=0.23\textwidth]{(c)}{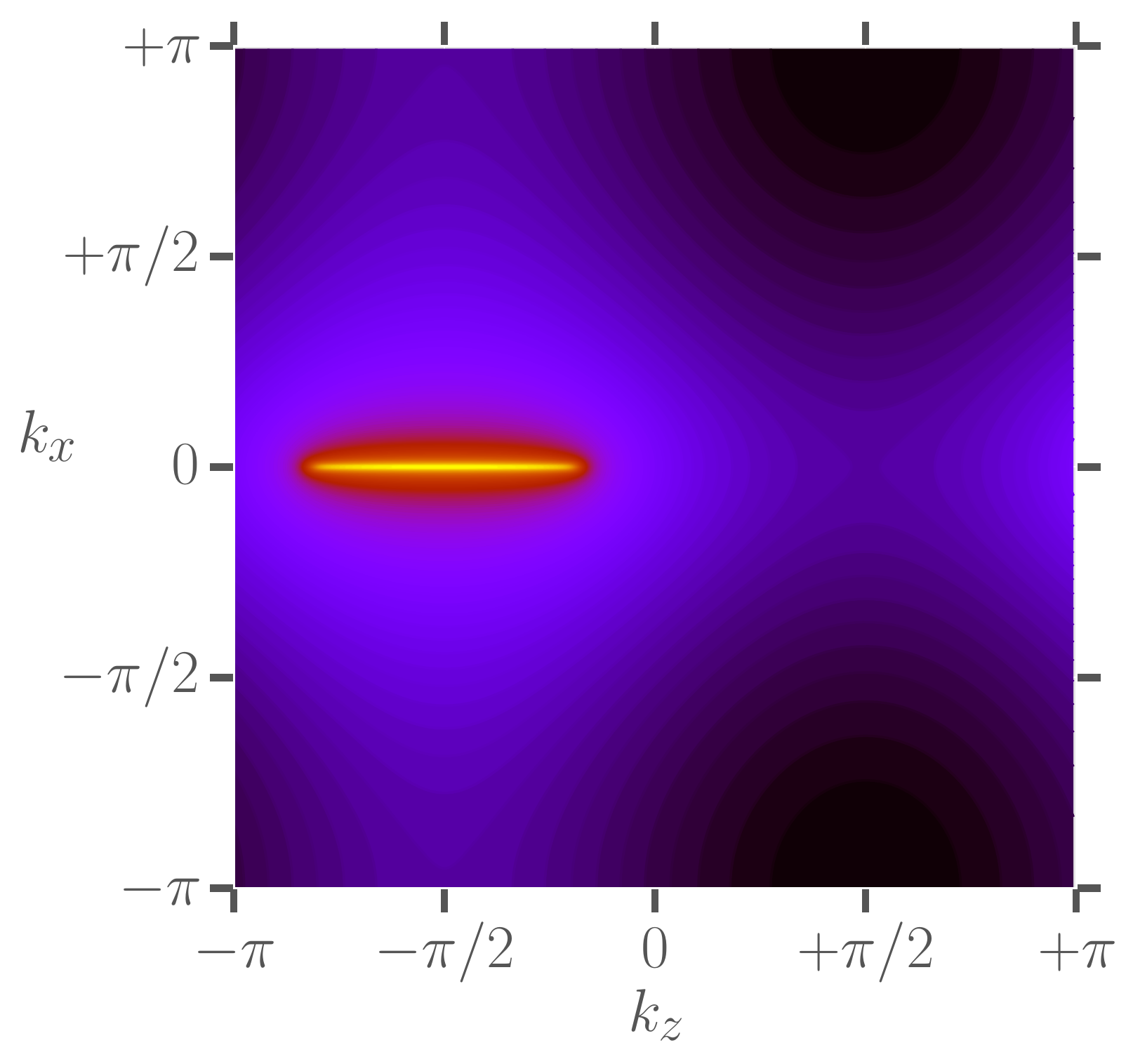}}\\[-0.3em]%\hspace*{-0.9em}
\subfloat{\subfigimg[width=0.23\textwidth]{(d)}{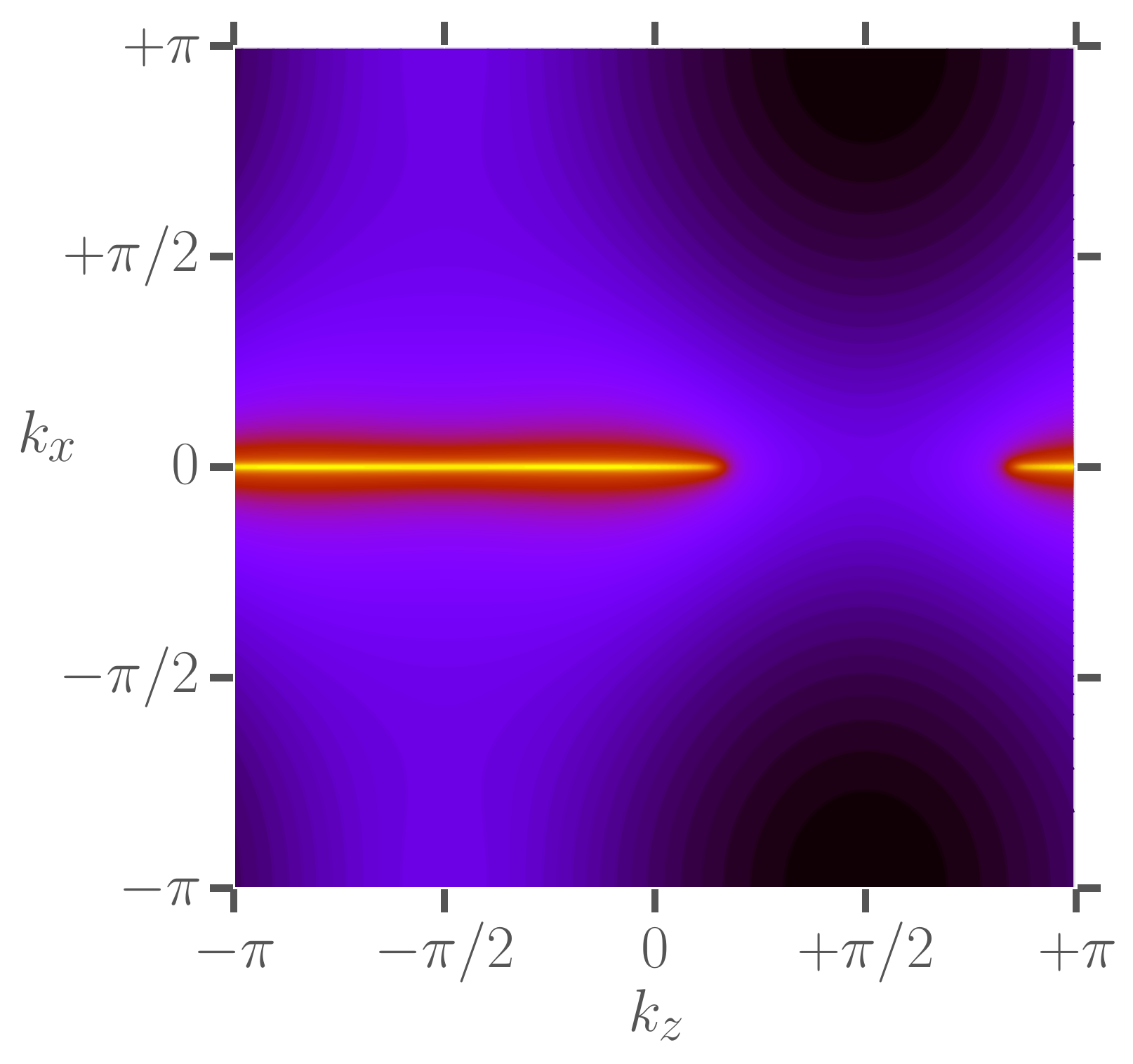}}
\caption{Numerical evaluation of the surface spectral function in Eq.~\ref{geff} (a) Surface spectral function as a function of energy $E=\omega$ and momentum $k_x$ for fixed momentum $k_z=-\frac{\pi}{2}$. (b) Surface spectral function as a function of momentum $k_z$ at energy $\omega=0$ and $\eta=0.01$ for varying arc lengths $m=0.9$, (c) $m=0.5$, (d) $m=-0.5$}\label{fig:surface_spectral}
\end{center}
\end{figure}

For realistic system sizes, the matrices involved are very large and the solution to Eq.~(\ref{geff}) is only accessible numerically. We perform the numerical calculation and present the spectral function $A = -\frac{1}{\pi} \Im{[\text{Tr}(G_{\text{eff}})]}$ in Fig.~\ref{fig:surface_spectral}, Panel~\ref{fig:surface_spectral} (a) shows the spectral function for the combined top and bottom surfaces for fixed $k_z$.  It is plotted as a function of $k_x$ and the energy. One can see the top and bottom surface states in the gap.  The positive slop line represents surface states on the top surface while the negative slope line contains states confined to the bottom surface.  These states exist for any $k_z$ between the two Weyl points and at zero energy they form the Fermi arcs. The Fermi arc states are also seen in the spectral function cut at zero energy, as seen in Figs~\ref{fig:surface_spectral} (b-d) for different arc lengths, controlled by the parameter $m$.

\subsection{Analytic Green's function at low energy, semi-infinite sample}
In order to advance analytically, we change the geometry of the system to a semi-infinite slab in $y$-direction by taking $N_y \rightarrow \infty$. Thus, we set $H_s = H_0$ and the only non-zero block of the matrix $T$ is equal to $R$. 
Looking at a semi-infinite slab has the advantage that the system with one layer removed is identical to the system before removing the layer.  We therefore envision that we're looking for the effective Green's function for the $n$th layer when the effective Green's function for the $(n+1)$th is known.  Since the system is unchanged by removing a single layer, the two Green's functions above are identical.  This leads to the following recursive equation:
\begin{align}
\begin{split}
G_{\text{eff}} = \left[ G_0^{-1}(i\omega_n) - R^\dagger G_{\text{eff}}R \right]^{-1}, \label{SC_equation}
\end{split}
\end{align}
where we have defined the uncoupled Green's function, $G_{0} = (i\omega_n - H_0)^{-1}$. The equation for $G_{\text{eff}}$ is now simply a $2\times2$ matrix equation and can be solved analytically. Nonetheless, for general parameter values, the solution is quite complicated and not very insightful.  It is therefore useful to simplify it by transforming the system via the unitary transformation $U = \exp{\left( -i\frac{\pi}{4} \sigma_y \right)}$, which corresponds to a rotation around the $y$-axis in orbital space. In addition, we set $t_s=t^\prime=1$ in order to simplify the result. This leads to,
\begin{align}
\begin{split}
&R = \begin{pmatrix} 0 & 0 \\ t_s & 0 \end{pmatrix}, \\
&H_0 =  g_1({\bf k})\sigma_z - g_3({\bf k}) \sigma_x.
\end{split}
\end{align}
The solution for the Green's function reads,
\begin{align}
\begin{split}
G(i\omega_n,k) = \begin{pmatrix}
G_{\text{eff}}^{(2)} & G_{\text{eff}}^{(3)} \\
G_{\text{eff}}^{(3)} & G_{\text{eff}}^{(1)}
\end{pmatrix},
\end{split}
\end{align}
where
\begin{align}
\begin{split}
G_{\text{eff}}^{(1)} &= \frac{1}{2t_s^2(i\omega + g_1)}\left( t_s^2 + (i\omega)^2 -g_1^2 - g_3^2 \pm \sqrt{p}  \right), \\
G_{\text{eff}}^{(2)} &= \frac{i\omega + g_1}{2t_s^2g_3^2}\left( -t_s^2 + (i\omega)^2 -g_1^2 - g_3^2 \pm \sqrt{p}  \right), \\
G_{\text{eff}}^{(3)} &= \frac{1}{2t_s^2 g_3}\left( t_s^2 - (i\omega)^2 +g_1^2 + g_3^2 \mp \sqrt{p}  \right). \label{solution}
\end{split}
\end{align}
and
\begin{align}
\begin{split}
p =& -4t_s^2(i\omega - g_1)(i\omega + g_1) + (-t_s^2-(i\omega)^2 + g_1^2 + g_3^2)^2.
\end{split}
\end{align}
Analytic continuation then yields the retarded/advanced Green's functions. The low energy part of the spectrum is governed by the poles of $G^{(1)}$ at $\omega = -g_1({\bf k})$, where $G^{(2)}$ and $G^{(3)}$ approximately vanish for on-shell momenta. In this regime we find:
\begin{align}
\begin{split}
G_{\text{eff,ret}}^{(1)} = \frac{t_s^2-g_3^2 + |t_s^2-g_3^2|}{2t_s^2\left( \omega +\sin{k_x} + i\eta \right)},
\end{split}
\end{align}
which leads to the spectral function
\begin{align}
\begin{split}
A_{\text{eff}}^{(1)} \propto \begin{cases}
(1 - \frac{g_3^2}{t_s^2})\delta(\omega + \sin{k_x}) &\text{for $\frac{g_3^2}{t_s^2}<1$}\\ \label{spectral}
0 &\text{otherwise}
\end{cases}
\end{split}
\end{align}
Therefore, at low energy the weight of the spectral function is concentrated in a limited part of the Brillouin zone. The zero energy states are obtained from the Green's function by setting $\omega=0$ and therefore $k_x=0$ (we ignore the case of $k_x=\pi$ as there is vanishing spectral weight there).  Since we have set $t_s=t'=1$, this gives the condition:
\begin{align}
\begin{split}
|m+1+\sin{k_z}| < 1 \quad \Rightarrow \quad m + \sin{k_z} < 0.
\end{split}
\end{align}
The left hand side of the above expression is zero when $k_z$ is at the Weyl points and is negative when $k_z$ is between them.  Therefore the zero energy states reside on a straight line between the two Weyl point projections on the surface Brillouin zone.

\section{Application of a magnetic field}\label{Sec:field}
When put in magnetic field the density of states of metals oscillates as a function of inverse field.  The oscillation frequency is proportional to the area enclosed by the Fermi surface cross section.  This effect has been long utilized for characterization of materials.  In two dimensions these quantum oscillations represent the Landau level quantization and in three dimensions the levels broaden and even overlap due to the dispersion along the field direction.  Semiclassically the the oscillations can be viewed as follows.  In the presence of magnetic field the quasiparticles encircle the Fermi surface due to the Lorentz force in momentum space.  This closed orbit produces a maximum in the density of states when the quasiparticle phase accumulated during the motion is an integer times $2\pi$.  

In a Weyl semimetal the Weyl points give rise to bulk quantum oscillations in any field direction.  Importantly, the broadened Landau level include a zero energy level for each Weyl point which disperses linearly with the momentum along the field.  We refer to this level as the chiral level and later comment about possible gapping of this level due to inter-nodal scattering.   On the other hand, the surface states do not exhibit closed Fermi surfaces and quantum oscillations do not appear in the usual way.  Instead, a path which includes the arcs on the top and bottom surfaces as well as tunneling through the low energy bulk states was proposed by Potter {\it et al.}\cite{potter}  This semiclassical argument leads to surface level quantization of the form found in Eq.~\ref{arcLL}.

These semiclassical orbits require the use of low energy bulk modes near the Weyl points and therefore the bulk chiral Landau level is replaced by the fine quantization above.  Bulk like levels are broader and begin at energies corresponding to $n>1$ bulk Landau level.  Their quantization condition is different from that of the surface.  We discuss both types of oscillations in this section. 

\subsection{Bulk Landau levels of a WSM}
The bulk Landau levels can be obtained by considering the continuum low energy Weyl Hamiltonian, Eq.~(\ref{h_weyl}). In our case $\vv = 0$ and $h_{ij}$ is a diagonal matrix.  Therefore,
\begin{align}
\begin{split}
H = v_F k_x \sigma_x + v_F k_y \sigma_y + \chi v_z k_z\sigma_z,
\end{split}
\end{align}
where  $v_z =a\sqrt{t_s^2 - m^2}$, $v_F = t_s a$ and $\chi$ is the chirality of the node. Here, $a$ is the lattice constant. We apply a magnetic field in the $y$-direction via substituting the canonical momentum $\mathbf{\pi} = \kv + {e\over \hbar c}\mathbf{A}$ and choosing the Landau gauge, $\mathbf{A}=-Bxe_z$. Defining raising and lowering operators gives:
\begin{align}
\begin{split}
v_z \pi_z = -i\sqrt{v_Fv_z\over 2 \ell_B^2}(a^\dagger - a), \quad v_F\pi_x = \sqrt{v_Fv_z\over 2 \ell_B^2}(a^\dagger + a).
\end{split}
\end{align}
where $\ell_B = \sqrt{\hbar c / eB}$ is the magnetic length. 
The spectrum is found by squaring the Hamiltonian:
\begin{align}
\begin{split}
H^2 = \left(\frac{2v_Fv_z}{\ell_B^2}(a^\dagger a + \frac{1}{2}) +v_F^2k_y^2 \right)\sigma_0 -\chi \frac{v_Fv_z}{\ell_B^2}\sigma_y,
\end{split}
\end{align}
where we set $\hbar=1$ from now on. Clearly the eigenstates of $H$ are eigenstates of $\sigma_y$. With the ansatz $\psi_0 = (|0\rangle, i\chi|0\rangle)^T$, the zeroth Landau level dispersion is found to be $E_0 = \chi v_Fk_y$. Thus, we end up with a single chiral Landau level, independent of the magnetic field. Higher Landau levels are given by
\begin{align}
\begin{split}
E_n = \pm \sqrt{\frac{2v_Fv_z}{\ell_B^2}n+v_F^2k_y^2}.
\end{split}
\end{align}
Note that each Weyl node exhibits only one zeroth Landau level. As explained by the Nielsen-Ninomiya theorem\cite{nielsen2,friedan}, Weyl nodes always come in pairs with opposite chiralities and in the full lattice model the chiral Landau levels are connected at high energy. 

\subsection{Surface quantum oscillations}
\subsubsection{Frequency}
We now turn to the quantum oscillations associated with the surface Fermi arcs and analyze them using our effective surface theory.  This allows us to test the semiclassical quantization condition of Eq.~\ref{arcLL} and extend it beyond the decoupled Weyl points regime.  The decoupled Weyl point approximation is valid for arc length $k_a$ such that   $1/k_a  \gg \ell_B$ and we therefore calculate the spectrum in and out of this regime. This is done by numerically solving the effective Green's function equation \ref{SC_equation}.

\begin{figure}[t]
\begin{center}
\subfloat[]{\includegraphics[width=0.25\textwidth]{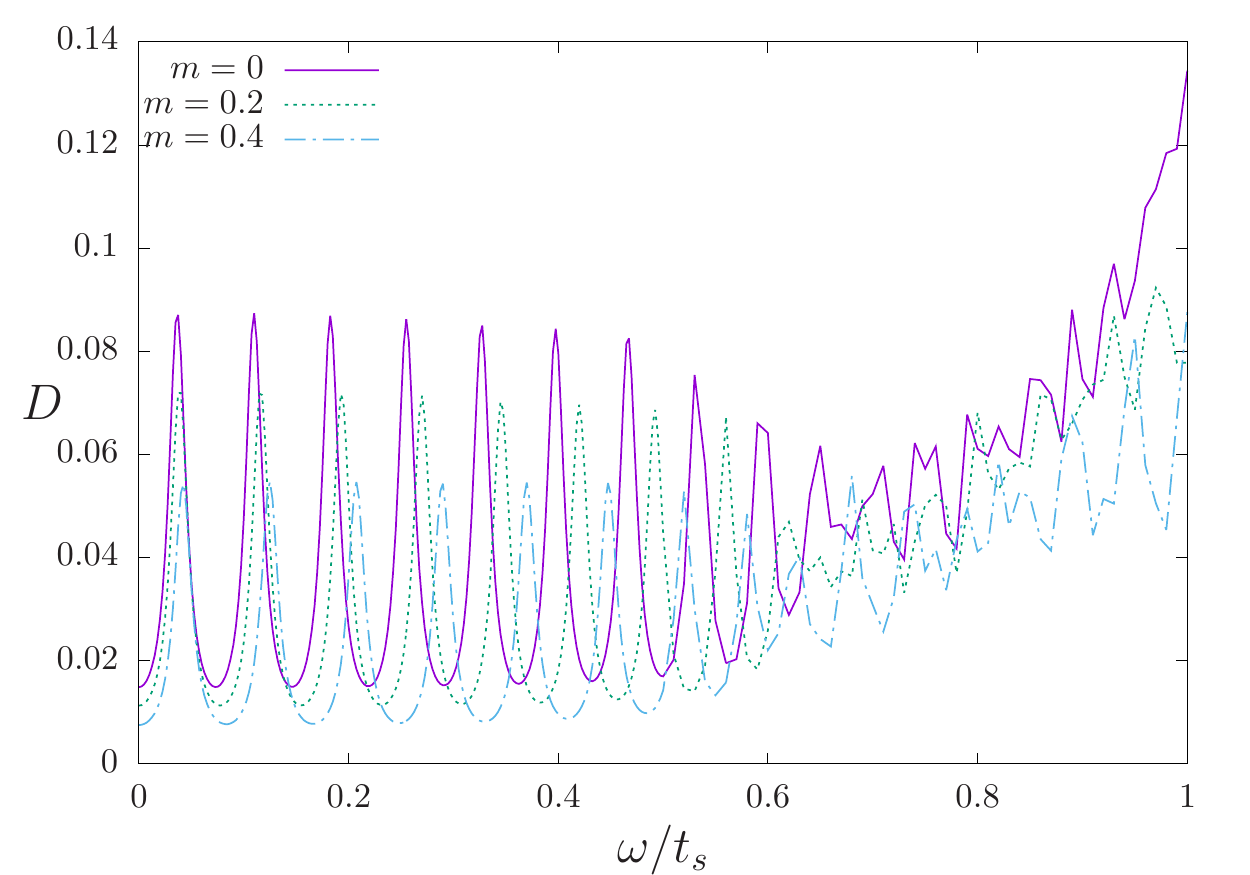}}\hspace*{-0.9em}
\subfloat[]{\includegraphics[width=0.25\textwidth]{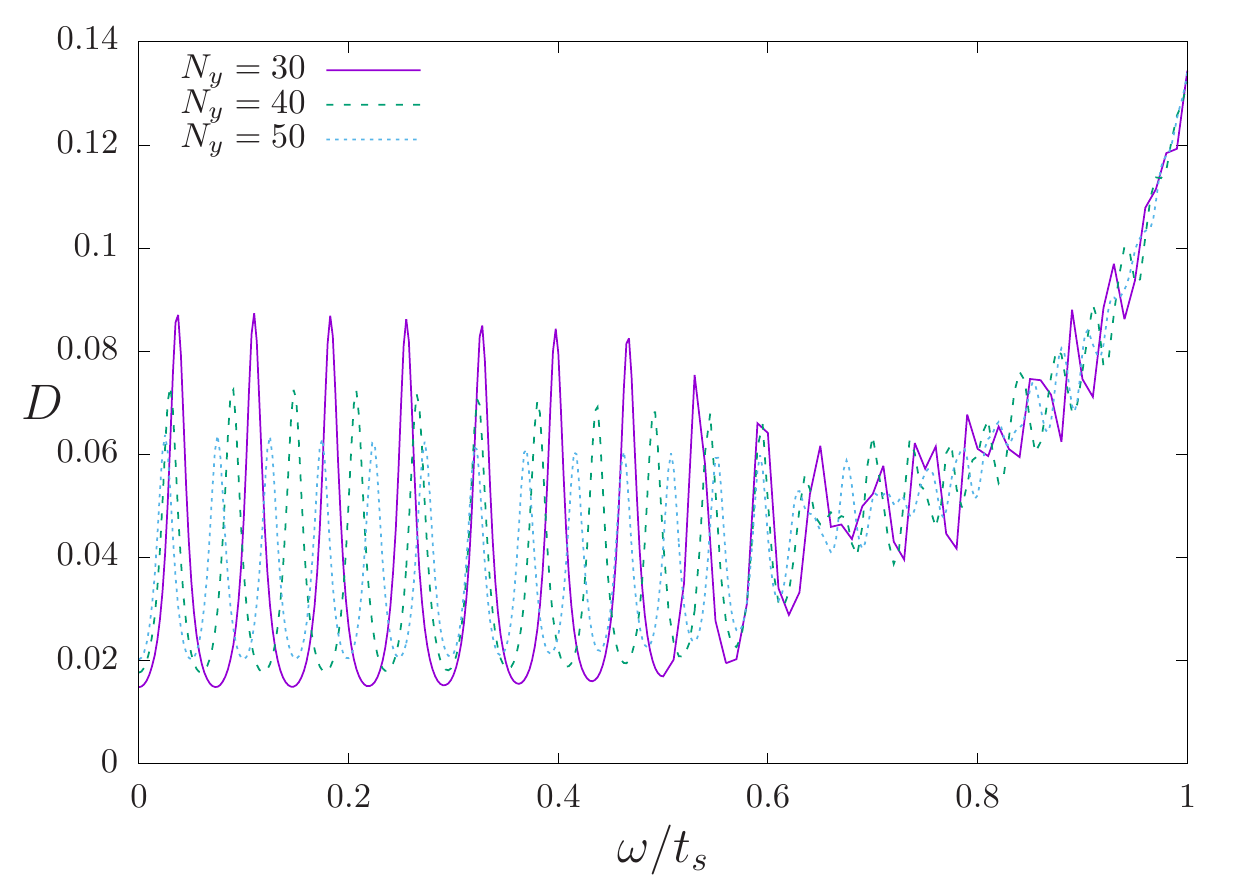}}
\caption{Surface density of states for (a) varying arc length at magnetic field with $q=30$ and thickness $N_y = 30$, (b) varying slab thickness with $q=30$ and $m=0$.
}\label{fig:surface_dos}
\end{center}
\end{figure}

In Fig.~\ref{fig:surface_dos} (a) we show the surface density of states (DOS) calculated using the effective Greens function Eq.~(\ref{geff}) for varying arc lengths in the long arc length regime for a fixed magnetic field and fixed slab thickness. 
In these graphs we represent the magnetic field by $q$, the number of unit cells in a magnetic unit cell.  The magnetic unit cell is chosen such that the flux threading it is the flux quantum $\Phi_0$.  
This analysis produces Landau level separation which is in good agreement with Eq.~\ref{arcLL}. In order to further test the compatibility of the effective surface model with the semiclassical theory we repeat this analysis in Fig.~\ref{fig:surface_dos}(b) for varying slab thickness. In these cases we find that the observed oscillations and the effective surface theory coincide well with the semiclassical theory.  For higher energies the clear oscillatory behaviour gets obscured by the fact that the bulk Landau levels overlap with the Fermi arc oscillations.
 
\subsubsection{Phase offset}
Another physical quantity that can be extracted from the plots is the phase offset $\gamma$.  The graphs suggest that there is no Berry phase contribution to the surface Landau levels as $\gamma=\frac{1}{2}$.  This can be interpreted as the cancellation of the Berry phase contributions of the two Weyl nodes. This is consistent with both chiral Landau levels participating in producing the surface Landau levels.

\begin{figure}[t]
\begin{center}
\subfloat[]{\includegraphics[width=0.25\textwidth]{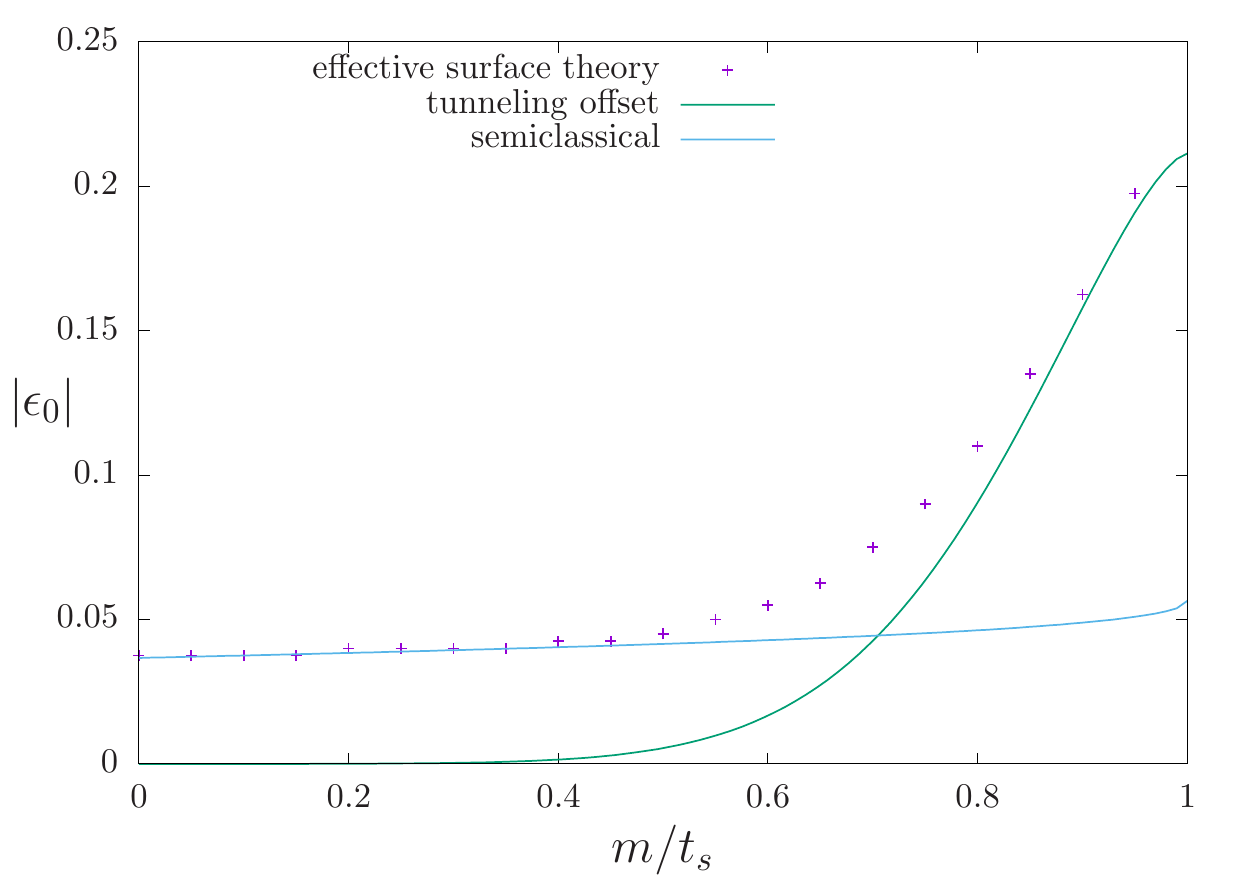}}\hspace*{-0.9em}
\subfloat[]{\includegraphics[width=0.25\textwidth]{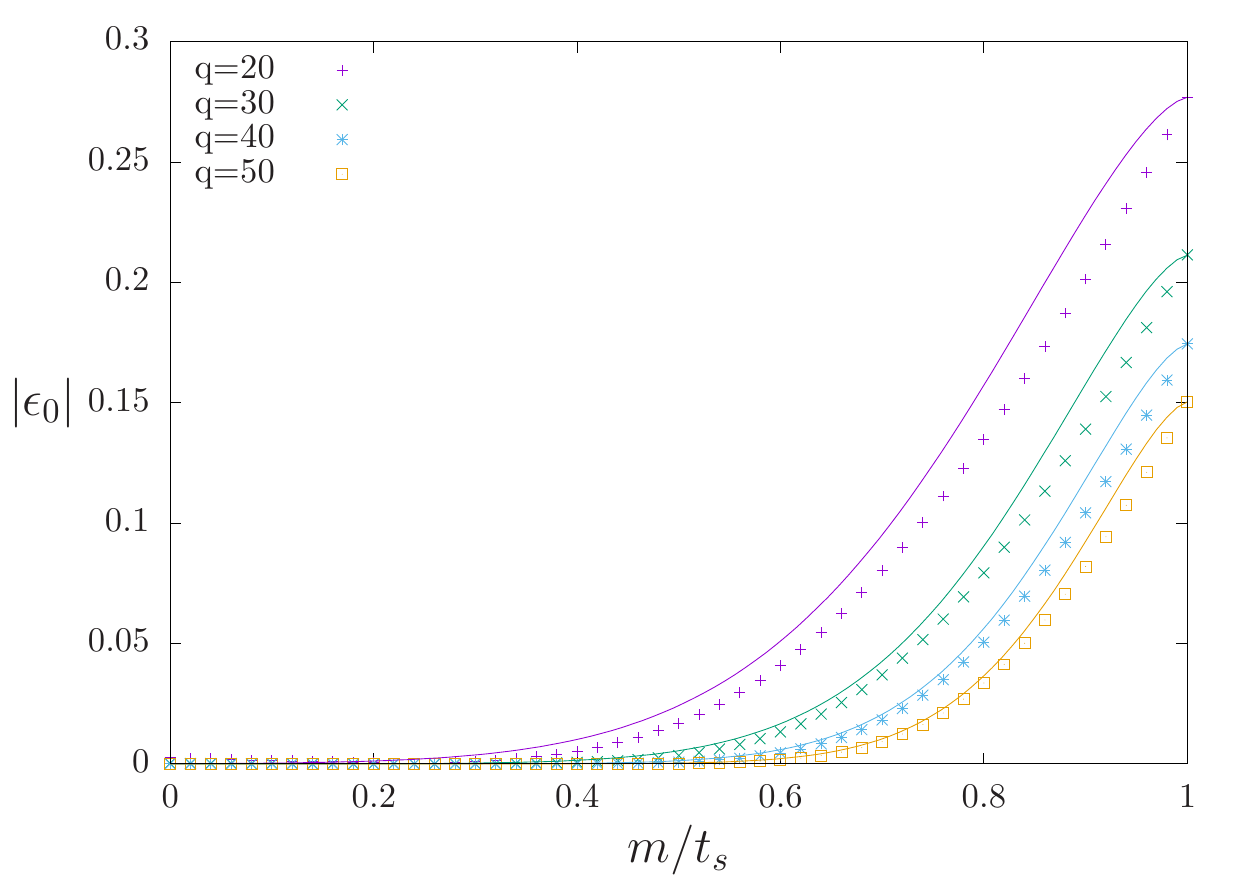}}\hspace*{-0.9em}
\end{center}
\caption{(a) Energy offset of the zeroth \emph{surface} Landau level for varying arc lengths. (b) The energy for the $n=0$, chiral \emph{bulk} Landau level as a function of $m$ as calculated numerically from the lattice model (data points) and in the WKB approximation in Eqs.~(\ref{eq:energyoffset1}-\ref{gap_main}) (lines) for varying magnetic flux $Ba^2=\frac{\Phi_0}{q}$, which is measured in flux quanta per $q$ unit cells.
}\label{fig:LL1}
\end{figure}

Moving from long arc lengths limit towards the small arc length regime, we expect hybridization between the two chiral Landau levels. This results in a gap and therefore a contribution to the phase offset. Indeed, when analyzing the energy offset in Fig.~\ref{fig:LL1} (a) one can see that for short arc lengths a gap opens between the positive and negative surface Landau levels and the quantum oscillations vanish when approaching the point where the two Weyl nodes fuse at $m=t_s$. 
We find that the offset $\gamma$ in the short arc length regime originates from gapping of the chiral Landau level due to Weyl point mixing.  We estimate the hybridization energy and compare it with the surface Landau level spectrum gap.  This can be seen in Fig.~\ref{fig:LL1}. 

In the short arc length limit the hybridization between two chiral Landau levels can be analyzed as tunneling in a double well potential in momentum space. In the appendix we analyze this problem using the WKB approximation and find the following energy splitting:
\begin{align}
\begin{split}\label{eq:energyoffset1}
\Delta \epsilon = \sqrt{2} C (m^*v_F^2\hbar^2\omega_c^2)^{\frac{1}{3}}\exp{\left( - \frac{2}{3} \left(\frac{k_a a}{2}\right)^3\frac{m^*v_F^2}{\hbar\omega_c} \right)},
\end{split}
\end{align}
where
\begin{align}
\begin{split}
C =  \frac{1}{\sqrt{2}} \left( 2 \sqrt{\frac{\pi}{2}}\frac{\Gamma(\frac{7}{4})}{\Gamma(\frac{1}{4})} \right)^{\frac{2}{3}} \approx 0.523\label{gap_main}
\end{split}
\end{align}
where we have explicitly included the lattice constant $a$. The Fermi velocity $v_F$, the effective mass $m^*$ for our model are defined in the appendix and $\omega_c$ is the cyclotron frequency. This offset is derived from the bulk model and in Fig.~\ref{fig:LL1} (b) it is compared to the value of the zeroth Landau level at $k_y=0$ for the full bulk lattice model. One can see that the construction overestimates the energy gap by a small amount. This is due to the fact that the WKB approximation used in the derivation works better for higher Landau levels and the fact that we ignored a linear term in the potential. Nonetheless, the approximation captures the behaviour of the full system well.

Another interesting regime is a type II Weyl semimetal.  In this regime, due to the tilt term there is no chiral Landau level while other bulk Landau levels are present. As suggested by the semiclassical analysis, the existence of surface Landau levels depends crucially on the bulk chiral Landau level\cite{soluyanov}.  In its absence we do not expect to see surface level quantization.

\begin{figure}[t]
\begin{center}
\subfloat[]{\includegraphics[width=0.5\textwidth]{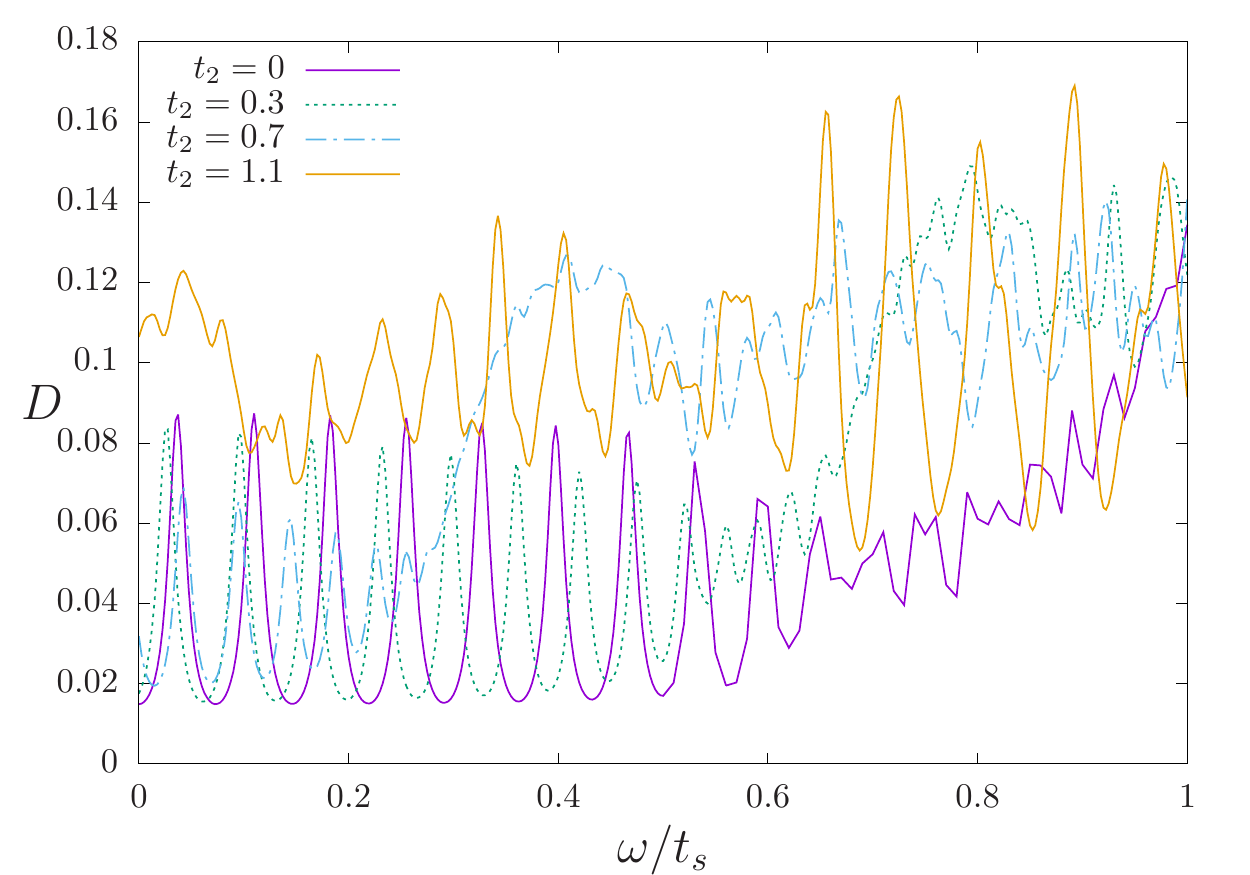}}
\end{center}
\caption{ Surface density of states for $q=30, N_y=30$ and $m=0$ for varying parameter $t_2$, measured in units of $t_s$.}\label{fig:LL}
\end{figure}

To test whether the surface states are quantized in Landau levels in a type II WSM we add a term to the lattice Hamiltonian in Eq.~\ref{eq:H3D}
\begin{equation}
H_2 = t_2 \sin(k_z),
\end{equation}
which turns the Weyl nodes into type-II Weyl nodes for $t_2 >t_s$. In this regime $H_2$ completely dominates the spectrum and applying a magnetic field in a direction perpendicular to $z$ leads to a gapped spectrum. In Fig.~\ref{fig:LL} we show the results for various values of $t_2$, where one can see that when increasing $t_2$ the low energy regime which is dominated by the Fermi arc quantum oscillations shrinks until it completely vanishes for $t_2>t_s$.  Our results therefore support the claim that the chiral Landau level is necessary for the formation of surface Landau levels.

\section{Three Dimensional Lattice Model Analysis}\label{Sec:numeric}
In order to test the predictions of our effective surface theory, we use numerical diagonalization of a lattice model with an applied magnetic field via Peierls substitution. As in previous sections, the magnetic field $B$ is oriented in the $y$-direction and in the Landau gauge. In this gauge the hopping along $z$ acquires an $x$-dependent phase which breaks the translation invariance in the $x$-direction.  We therefore define a magnetic unit cell, elongated along the $x$-direction.  Choosing a cell of length $q$ lattice constants through which a flux quantum $\Phi_0$ is threaded amounts to a magnetic field $B = \Phi_0/qa^2$ where $a$ is the lattice constant.  We vary $q$ to control the field strength.  With this gauge the hopping along $z$ acquires a phase of $\exp\left(-i\frac{2\pi n_x}{q}\right)$, where $n_x$ is the index of the $n$th lattice site inside the magnetic unit cell. This increases the sizes of the matrices $H_0$ and $R$ to $2q$-by-$2q$ and the $y$-layer Hamiltonian reads
\begin{align}
\begin{split}
H_0 = &\sum_{k_x,k_z}\left[\sum_{n=1}^q (m+2+\sin{\left( kz-2\pi/q\cdot n \right)} )\sigma_z c^\dagger_n c_n \right. \\
&\left. +  \sum_{n=1}^{q-1}( \frac{i\sigma_x-\sigma_z}{2})c^\dagger_n c_{n+1}  \right.\\
& \left.-\frac{1}{2} e^{-ik_x}\sigma_z c^\dagger_1 c_q + \frac{1}{2i}e^{-ik_x}\sigma_xc^\dagger_1c_q+ \text{h.c.}\right] ,
\end{split}
\end{align}
where we have suppressed the $k$-indices on the creation/annihilation operators. The coupling between different $y$-layers is given by
\begin{align}
\begin{split}
R = \left( -\frac{1}{2}\sigma_z + \frac{1}{2i}\sigma_y \right) \otimes {\rm 1}_q,
\end{split}
\end{align}
where ${\rm 1}_{q}$ is a $q$-by-$q$ unit matrix in the magnetic unit cell basis. The full $2qN_y$-by-$2qN_y$-Hamiltonian can be easily constructed. When diagonalizing the full system, we expect the low energy spectrum to be dominated by the surface quantum oscillations and by bulk contributions at higher energies.
\begin{figure*}[t]
\begin{center}
\subfloat[~]{\includegraphics[width=0.45\textwidth]{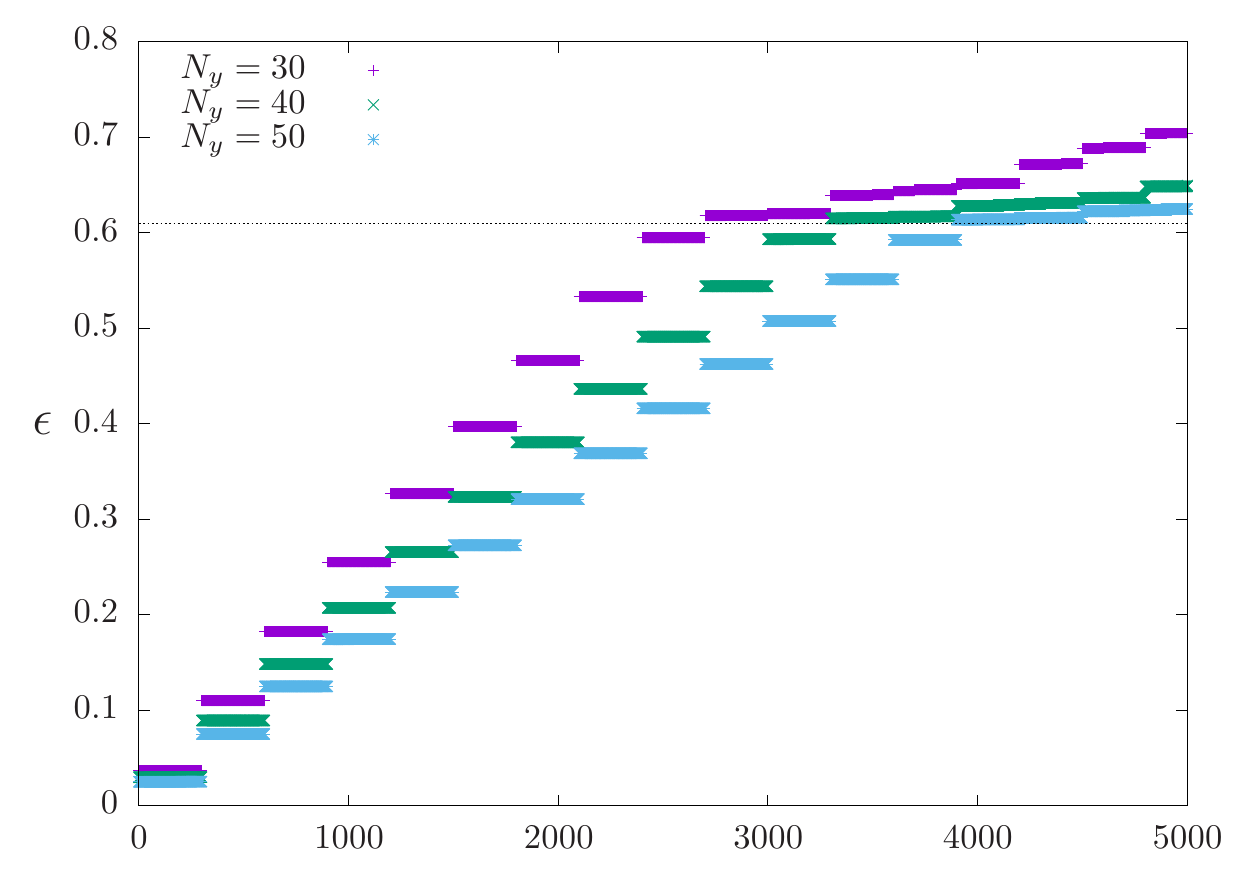}}
\subfloat[~]{\includegraphics[width=0.45\textwidth]{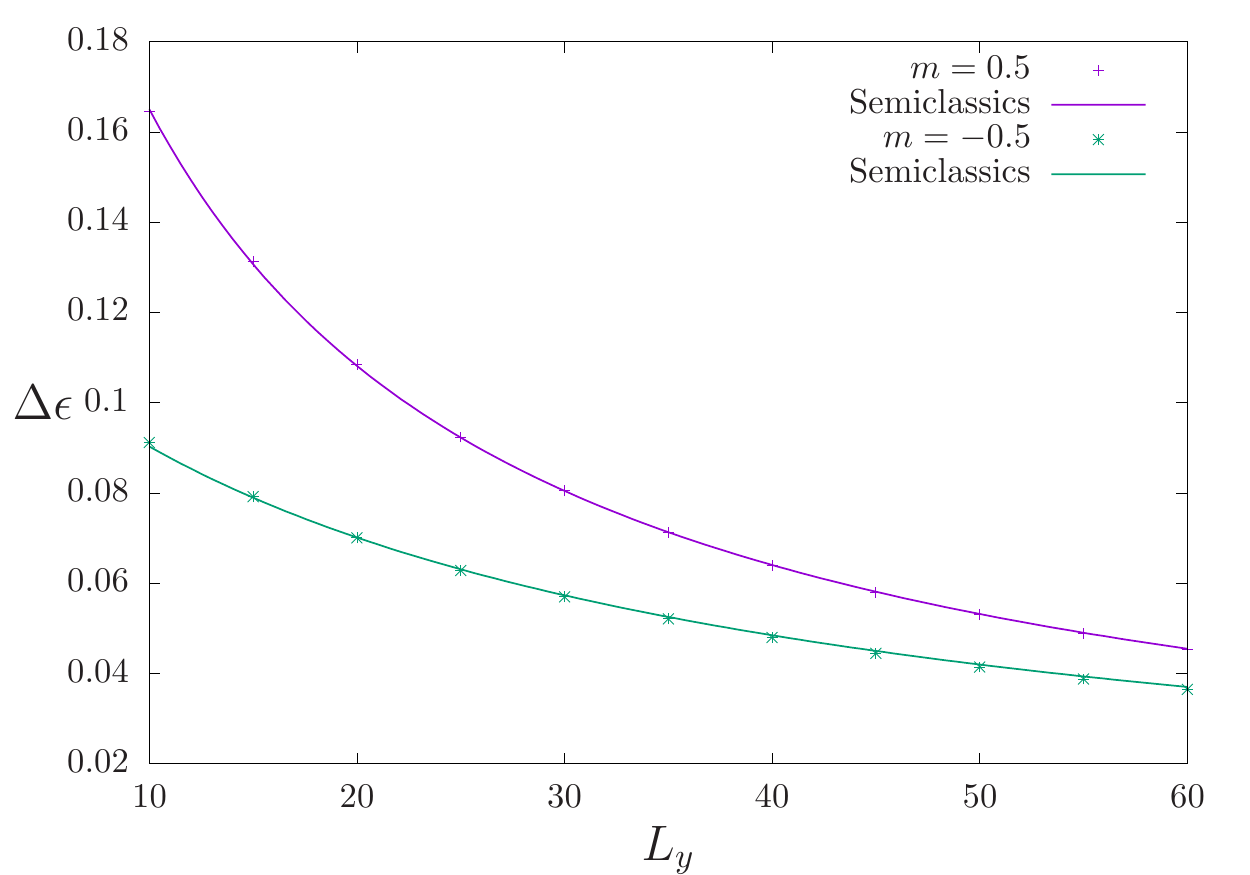}} \\
\subfloat[~]{\includegraphics[width=0.45\textwidth]{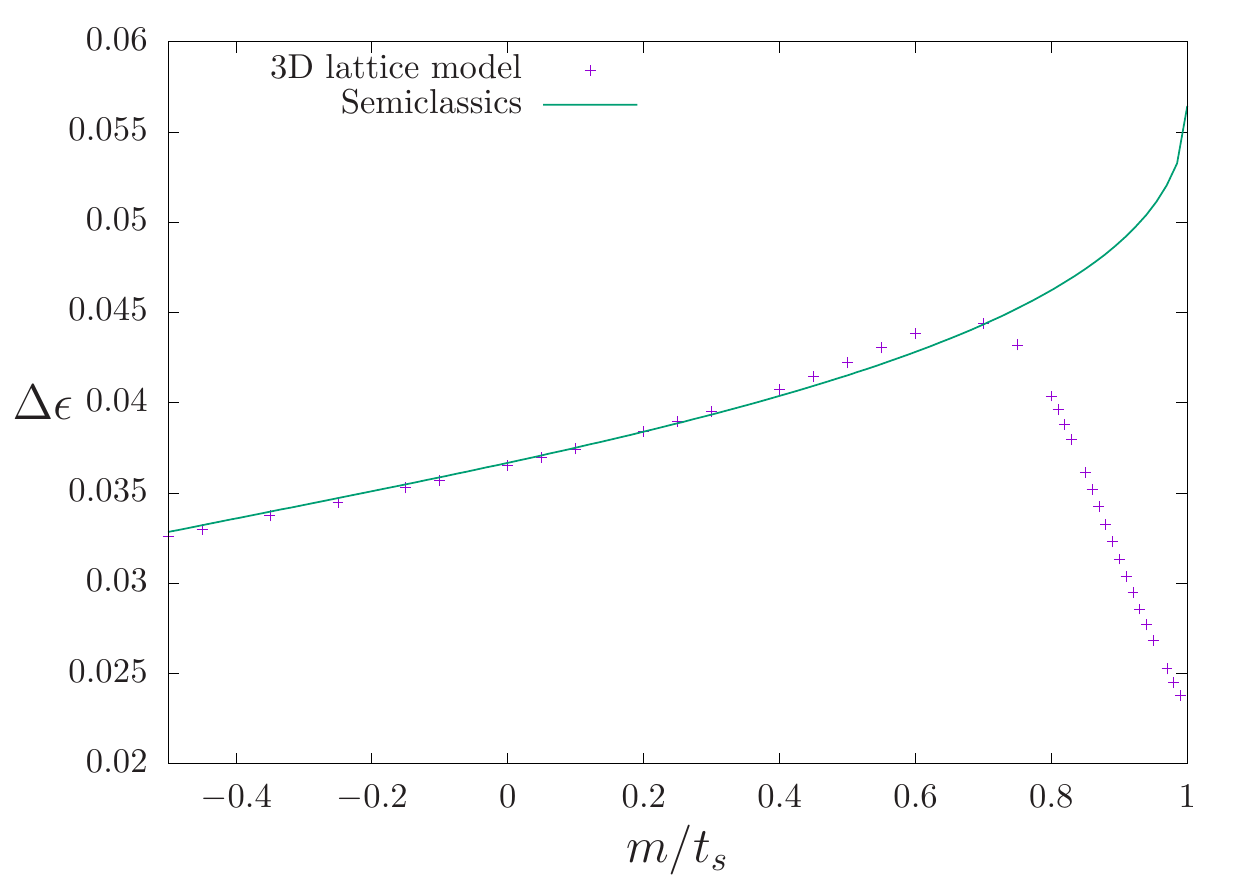}}
\subfloat[~]{\includegraphics[width=0.45\textwidth]{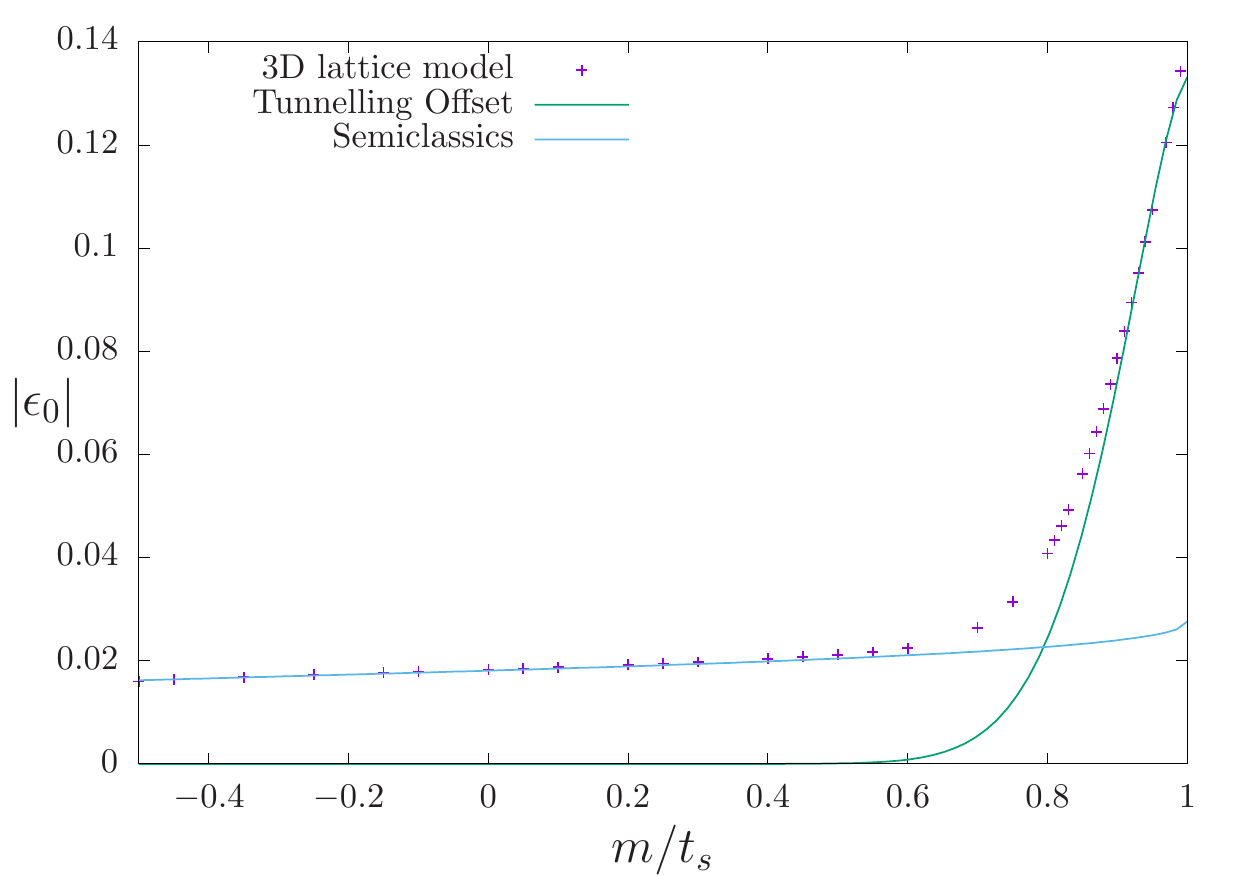}}
\caption{(a) Full 3D spectrum for a slab with model parameters $q=80$ and $m=0$ for varying slab widths. (b) Energy difference of the first and zeroth Landau level for $q=40$ as a function of slab width $N_y$. (c) Energy difference of the first and zeroth Landau level for $q=60$ and $N_y=60$ as predicted by the semiclassical theory (line) and the full 3D model. (d) Energy offset of the zeroth Landau level of the full 3D model as a function of arc length for the same parameter values as in (c). }\label{fig:bulk}
\end{center}
\end{figure*}

In Fig.~\ref{fig:bulk} (a) we show the low energy spectrum of a WSM slab in magnetic field.
At low energy (below $0.6 t_s$ in our model) one can see the Fermi-arc Landau level structure.
For higher energies, one can observe bulk Landau levels mixed with the surface Landau levels. This is shown explicitly by varying the slab thickness. The frequency of the Fermi arc induced quantum oscillations varies with slab thickness, but the bulk Landau level spacing is approximately constant. 

Analyzing the frequency of the the Landau levels enables us to test the predictions of the semiclassical theory and in addition further investigate the behaviour of the slab in regimes not accessible by the semiclassical theory. This is done in Fig.~\ref{fig:bulk}(b) where we confirm that the semiclassical approach accurately describes the frequency of the quantum oscillations when varying the slab width. Further, in Fig.~\ref{fig:bulk}(c) we vary the arc length $k_a = 2\arccos{(\frac{m}{t_s})}$ by varying $m/t_s$ and compare the energy difference of the first and zeroth Landau level with the semiclassical theory Eq.~(\ref{arcLL}). One can see that only for long arc lengths does the semiclassical theory describe the full quantum model well. This was done by taking into account the diabatic correction\cite{potter} to the arc length $\propto \ell_b^{-1}$ which is due to the fact that an electron on the arc can tunnel through the bulk even before reaching a Weyl point. For small arc lengths the behaviour significantly deviates from the semiclassical theory and converges towards the behaviour of the merged Weyl points derived in appendix \ref{bulkLL}.

Further, we can analyze the phase offset through the full bulk system and compare it to the findings of the effective surface theory. In Fig.~\ref{fig:bulk}(d) we show a comparison of the energy offset in the full bulk system with the semiclassical expectation with $\gamma = \frac{1}{2}$ as well as the tunnelling gap. One can see the expected crossover behaviour from the offset governed by the surface contribution to the bulk behaviour for $m\rightarrow t_s$.

\section{Conclusion}\label{Sec:conclusion}
In this work we derived an effective surface theory for the surface of a Weyl semimetal.  We used this theory as well as the full three dimensional lattice model to study the quantum oscillations.  We find that the surface theory accurately predicts the quantum oscillations which are associated with the surface states. Our model allows us to find the regime of validity of the semiclassical analysis of the surface quantum oscillations.  We find that it fails when the Weyl points hybridize.  This hybridization gaps the bulk chiral Landau level and in turn gaps the surface quantum oscillations which are essentially a fine quantization inside the chiral Landau level.  We estimate the hybridization gap using a double well potential model analyzed in the WKB approximation and find good agreement with the three dimensional model.

\appendix

\section{Bulk Landau levels for overlapping Weyl nodes} \label{bulkLL}
For large Weyl node separations, one can model each Weyl node separately and arrive at the usual Weyl Landau levels. However, when the two Weyl nodes approach each other, they can not be treated separately.  As a result the pair of chiral Landau levels hybridize and the spectrum developes a gap at zero energy.

In order to address this we consider a bulk model with an applied magnetic field in the $y$-direction, $\vec B = B\vec e_y$ and vector potential $\vec A = Bz \vec e_x$. As discussed in the introduction, the system experiences a gap closure at $m=t_s$. Lowering $m$ further, the gap closure splits into two Weyl nodes which traverse the Brillouin zone and recombine for $m=-t_s$. The point of the gap closure for $m=t_s$ is $\vec k_0 = (0,0,-\pi/2)^T$. Close to this point one can expand the Hamiltonian as
\begin{align}
\begin{split}
H = v_F k_x \sigma_x + v_F k_y \sigma_y + (\gamma+\frac{k_z^2}{2m^*})\sigma_z,
\end{split}
\end{align}
where $\gamma = m-t_s$ and we have omitted the quadratic terms in $k_x$ and $k_y$. We defined the Fermi velocity $v_F = ta$ and $m^* = \frac{1}{a^2 t_s}$, where we explicitly wrote out the lattice constant $a$. Adding the magnetic field to the system, we arrive at 
\begin{align}
\begin{split}
H = Bv_F\tilde z \sigma_x + v_F k_y \sigma_y + (\gamma+\frac{k_z^2}{2m^*})\sigma_z,
\end{split}
\end{align}
where we define $\tilde z = z + k_x/B$. Normally one arrives at a zeroth Landau level which depends on the momentum along the field direction, in this case $k_y$, and is gapless for $k_y=0$. Here, however, the zeroth Landau level acquires a mass which is estimated below.

Defining the new variables $Z = \frac{\tilde z}{\alpha}$ and $K = \alpha k_z$ with $\alpha = (v_F m^*B)^{-\frac{1}{3}}$, we arrive at
\begin{align}
\begin{split}
H = (m^*v_F^2\omega_c^2)^{\frac{1}{3}} \left( Z\sigma_x + (\Gamma + \frac{K^2}{2})\sigma_z \right),
\end{split}
\end{align}
with the now dimensionless $\Gamma = \frac{\gamma}{(m^*v_F^2\omega_c^2)^{\frac{1}{3}}}$ and the cyclotron frequency $\omega_c = \frac{B}{m^*}$. Squaring the Hamiltonian leads to
\begin{align}
\begin{split}
H^2 =& {}(m^*v_F^2\omega_c^2)^{\frac{2}{3}}
\begin{pmatrix}
Z^2 + (\Gamma + \frac{K^2}{2})^2 & -\frac{1}{2}[Z,K^2] \\
\frac{1}{2}[Z,K^2] & Z^2 + (\Gamma+ \frac{K^2}{2})^2
\end{pmatrix} \\
=&{}(m^*v_F^2\omega_c^2)^{\frac{2}{3}}\left[ \left( Z^2 + (\Gamma + \frac{K^2}{2})^2 \right) \sigma_0 + K\sigma_y\right], \label{squared}
\end{split}
\end{align}
where we have used $[Z,K^2] = 2iK$. Ignoring the prefactor for now, one gets the eigenvalue equation,
\begin{align}
\begin{split}
 \left( Z^2 + (\Gamma + \frac{K^2}{2})^2 \right) c_1 -iKc_2 = E^2c_1, \\
 iKc_1 +  \left( Z^2 + (\Gamma + \frac{K^2}{2})^2 \right)c_2 = E^2 c_2.
\end{split}
\end{align}
Using the ansatz $c_2 = ic_1$, we get
\begin{align}
\begin{split}
 \left( Z^2 + (\Gamma + \frac{K^2}{2})^2 + K \right)c_{1,2} = E^2c_{1,2}.
\end{split}
\end{align}
This has the general structure of the differential equation for an anharmonic oscillator when the operators are written in the {\it momentum space} basis as opposed to the real space basis. Using this analogy, we can solve this by defining the 'potential' in momentum space, $V \equiv V(K) = (\Gamma + \frac{K^2}{2})^2 + K$. In order to analyze the equation, we approximate it by ignoring the linear term. We want to investigate the $m=t_s$ limit, i.e., $\Gamma=0$. In this case the two Weyl nodes are combined to one single gap closure and one can solve the problem via the WKB approximation. The quantization condition reads,
\begin{align}
\begin{split}
\int_{x_-}^{x_+}\sqrt{E^2-V} = \left( n + \frac{1}{2} \right)\pi,
\end{split}
\end{align}
where $x_{\pm} = \pm \sqrt{2}E^{\frac{1}{2}}$ are the turning points. The left hand side can be transformed,
\begin{align}
\begin{split}
E\int_{x_-}^{x_+}dK \sqrt{1-\left(\frac{K}{x}\right)^4} = \frac{\sqrt{2}E^{\frac{3}{2}}}{2}\int_0^1 \frac{dt}{t^{\frac{3}{4}}}\sqrt{1-t},
\end{split}
\end{align}
with $t = \left(\frac{K}{x}\right)^4$. The integral is defined as the Euler beta function,
\begin{align}
\begin{split}
\mathcal{B}(x,y) = \int_0^1dt \ t^{x-1}\left( 1-t \right)^{y-1},
\end{split}
\end{align}
at $x=\frac{1}{4}$ and $y=\frac{3}{2}$ and therefore,
\begin{align}
\begin{split}
\pi \left( n + \frac{1}{2} \right) = \frac{\sqrt{2}E^{\frac{3}{2}}}{2}\mathcal{B}\left(\frac{1}{4},\frac{3}{2}\right).
\end{split}
\end{align}
With $\mathcal{B}\left(\frac{1}{4},\frac{3}{2}\right) = \frac{\sqrt{\pi}\Gamma(\frac{1}{4})}{2\Gamma(\frac{7}{4})}$ we get
\begin{align}
\begin{split}
E_n = \left( 4 \sqrt{\frac{\pi}{2}}\frac{\Gamma(\frac{7}{4})}{\Gamma(\frac{1}{4})}(n + \frac{1}{2}) \right)^{\frac{2}{3}},
\end{split}
\end{align}
and reinstating the prefactor leads to
\begin{align}
\begin{split}
E_n = (m^*v_F^2\omega_c^2)^{\frac{1}{3}}\left( 4 \sqrt{\frac{\pi}{2}}\frac{\Gamma(\frac{7}{4})}{\Gamma(\frac{1}{4})}(n + \frac{1}{2}) \right)^{\frac{2}{3}}.
\end{split}
\end{align}
As usual, the WKB approximation works better for higher Landau levels and we have ignored the linear term. When comparing to the numerically calculated Landau levels of the full system, only the first couple of Landau levels acquire a correction $\zeta(n)$. We find numerically that the zeroth Landau level gets a factor of $\zeta(0) \approx 0.811$ and already for $n\ge 1$ we find $\zeta(n) \approx 1$, which is very close to what the authors in [\onlinecite{dietl}] have found.
\section{Chiral Landau level for small Weyl node separation}
In order to estimate the energy shift of the chiral Landau level as a function of the Weyl node separation, we will start from equation (\ref{squared}), from which one can read off the potential in momentum space as $V(K) = (\Gamma+ \frac{K^2}{2})^2 + K$. This is the potential of an anharmonic oscillator in momentum space and for $\Gamma < 0$ we have two distinct minima. Ignoring the linear term, we have a symmetric double well problem\cite{landau}, where $\Gamma=\frac{m-t_s}{(m^*v_F^2\omega_c^2)^{\frac{1}{3}}}$ controls the separation of the two wells. We approach the problem by restricting our Hilbert space to that spanned by two wavefunctions, $\psi_r$ and $\psi_l$.  These are the ground states of the left and right wells located at $\pm\sqrt{2\Gamma}$, when they are completely separated form each other. When the two wells approach each other, the two wavefunctions hybridize due to the finite tunnelling probability. The new eigenstates are the symmetric and anti-symmetric combinations,
\begin{align}
\begin{split}
\psi_{\pm}(K) = \frac{1}{\sqrt{2}} \left ( \psi_r (K) \pm \psi_r(-K) \right).
\end{split}
\end{align}
The wavefunctions obey the Schr\" odinger equations,
\begin{align}
\begin{split}
\psi_r^{''} + (V-E_r)\psi_r = 0, \\
\psi_{\pm}^{''} + (V-E_{\pm})\psi_{\pm} = 0
\end{split}
\end{align}
where the first equation is valid due to the fact that the double well potential equals the single well potential in the regime of $\psi_r$. In addition, the amplitude of $\psi_r$ in the left well is vanishingly small.
Without loss of generality, we pick $\psi_+$. Multiplying the first equation by $\psi_+$ and the second one by $\psi_r$, subtracting the first equation from the second and subsequently integrating from $0$ to $\infty$ and using integration by parts, one arrives at,
\begin{align}
\begin{split}
\Delta E = 2\psi_r (0)\psi_r^{'} (0).
\end{split}
\end{align}
Restoring the prefactor from (\ref{squared}) and taking into account that we squared the Hamiltonian in order to derive Eq.~(\ref{squared}), via $\Delta \epsilon = (m^*v_s^2\omega_c^2)^{\frac{1}{3}}\sqrt{\Delta E}$  we arrive at,
\begin{align}
\begin{split}
\Delta \epsilon = \sqrt{2} (m^*v_F^2\omega_c^2)^{\frac{1}{3}} \left( \psi_r (0) \psi_r^{'} (0) \right)^{\frac{1}{2}}.
\end{split}
\end{align}
In order to evaluate this expression, we use the WKB approximation and we have,
\begin{align}
\begin{split}
\psi_r (0) = \frac{C}{V^{\frac{1}{4}}}\exp{\left( - \int_0^{\sqrt{2\Gamma}} |\sqrt{V} |\right)}, \quad \psi_r^{'}(0) = \sqrt{V}\psi_r(0)
\end{split}
\end{align}
Evaluating the integral we arrive at,
\begin{align}
\begin{split}
\Delta \epsilon &={} \sqrt{2} (m^*v_F^2\omega_c^2)^{\frac{1}{3}} C \exp{\left( -  |\Gamma|^{\frac{3}{2}} \left( \sqrt{2} + \frac{2^{\frac{3}{2}}}{6}\right) \right)}
\end{split}
\end{align}
The distance of the Weyl node from the middle point around which we have expanded the Hamiltonian, is given by $k_a/2 = \arccos(m/t_s) \approx \sqrt{2\frac{(t_s-m)}{t_s}}$, where we have expanded around $m=1$. With this, we arrive at the final result,
\begin{align}
\begin{split}
\Delta \epsilon = \sqrt{2} (m^*v_F^2\omega_c^2)^{\frac{1}{3}} C \exp{\left( - \frac{2}{3}\left(\frac{k_a}{2}\right)^3 \frac{m^* v_F^2}{\omega_c} \right)}. \label{gap}
\end{split}
\end{align}
The constant $C$ can be inferred from the lowest Landau level at $m=t_s$ (see Appendix~\ref{bulkLL}) and we get 
\begin{align}
\begin{split}
C =  \frac{1}{\sqrt{2}} \left( 2 \sqrt{\frac{\pi}{2}}\frac{\Gamma(\frac{7}{4})}{\Gamma(\frac{1}{4})} \right)^{\frac{2}{3}}.
\end{split}
\end{align}

\bibliographystyle{aipnum4-1}
\bibliography{weyl}

\end{document}